# Speaker Recognition using Deep Belief Networks to CCIS Proceedings

Adrish Banerjee, Akash Dubey, Abhishek Menon, Shubham Nanda,
Gora Chand Nandi,

Robotics and Artificial Intelligence Laborotary, Indian Institute of Information Technology,
Allahabad, India

**Abstract.** Short time spectral features such as mel frequency cepstral coefficients(MFCCs) have been previously deployed in state of the art speaker recognition systems, however lesser heed has been paid to short term spectral features that can be learned by generative learning models from speech signals. Higher dimensional encoders such as deep belief networks (DBNs) could improve performance in speaker recognition tasks by better modelling the statistical structure of sound waves. In this paper, we use short term spectral features learnt from the DBN augmented with MFCC features to perform the task of speaker recognition. Using our features, we achieved a recognition accuracy of 0.95 as compared to 0.90 when using standalone MFCC features on the ELSDSR dataset.

## 1  Introduction

Speaker recognition poses the major problem of understanding how to recognise and represent complex, high dimensional audio data. The solution used by most present day applications involve the use of mel frequency cepstral coefficients(MFCCs), highlighting the extensive dependence on domain specific feature engineering. We would therefore like to propose the use of unsupervised feature learning using a DBN[1] to represent the audio data in the form of spectral features.

Work by Ng et al. [2], Schwarz et al.[3] and Lei et al. [4] have shown significant improvement in the speech and speaker recognition tasks whilst deploying a phoneme based approach to learning. DBNs have also been known to work very effectively in the case of learning features from music audio [5], again capturing features based on the temporal aspects of data. But, keeping in mind the text independent nature of the problem, we have decided to approach the problem by learning hierarchical short term spectral features instead of hierarchical temporal features. Smith et al. [6] and Olshausen et al. [7] showed that sparse representation of audio signals were very similar to those of mammalian auditory processing neurons. For instance, the learned representations of natural sounds greatly resembled the cochlear filters in the auditory cortex. However, these methods have been used to learn shallow (1 layer) representations.

It is our hypothesis that learning more abstract representations of the speech signal would improve speaker recognition performance by producing better statistical models of the speech signal.These features, by encoding additional speaker dependent information, can hence be augmented with MFCCs to better model speaker characteristics. In this paper, we show that combining features learnt using DBNs with MFCCs produces better accuracy in speaker recognition systems.

## 2   Unsupervised Feature Learning using Deep Belief Networks

Deep belief networks are generative models with numerous layers of latent variables, which are typically binary. There is no intra-layer connectivity whereas the inter layer connections are undirected. Learning in these network turns out to be very difficult due to the intractability of inferring the posterior distribution from the hidden (latent) layers. Sampling from the posterior may be accomplished with deployment of MCMC (Markov Chain Monte Carlo) methods [8], however, these methods are extensively time consuming. [9] Presents a training procedure (using complementary priors to eliminate the explaining away effects) for the deep belief networks, which is equivalent to training a stack or sequence of Restricted Boltzmann Machines (RBM) as depicted in Fig. 1.

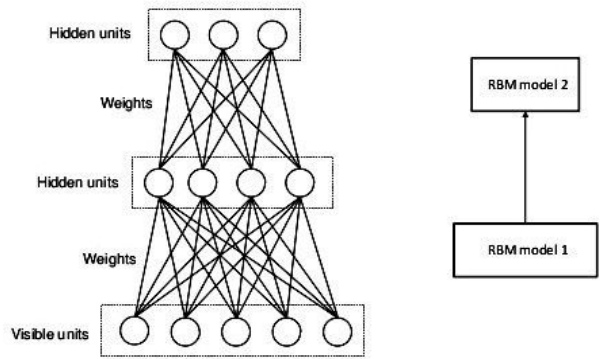

**Fig. 1.** Structure of the DBN used for extraction of short term spectral features, with two hidden layers, can be visualised as a stack of 2 RBMs.

### 2.1   Restricted Boltzmann Machines (RBM)

An RBM [10] (refer Fig. 2) is a type of Markov Random Field, with a layer of stochastic visible units and a layer of stochastic hidden units. There is no intra layer connectivity, but the all units of the visible layer are connected to all units of the hidden layer; the connections are weighted, undirected in nature. The weights of the connections along with the biases define a probability distribution over the state of the visible layer units. The energy function of a configuration of visible units *v* and hidden units *h* is given by,

$$E(v, h) = -b^T v - c^T h - v^T W h \tag{1}$$

Where E is the energy, *b* and *c* are the visible and hidden biases respectively and W is the weight matrix representing the strength of connections between visible and hidden units. The joint probability distribution and the conditional distribution of the j[th] unit of the hidden layer given the visible layer can be written as:

$$p(v, h) = \frac{1}{Z} e^{-E(v,h)} \tag{2}$$

$$p(h_j = 1|v) = sigmoid(c_j + v^T W_{:j})  \tag{3}$$

Where Z is the normalization or parition function and *:j* refers to all rows and the j[th] column of *W*.

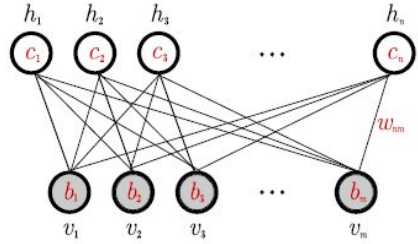

**Fig. 2.** Restricted Boltzmann Machine, where $v_1 .... v_m$ are visible units and $h_1 .... h_n$ are hidden (latent) units; $w_{ij}$ is the corresponding weight of the undirected connection between $v_i$ and $h_j$.

### 2.2 Generative Training of RBMs

The RBMs can be trained in a greedy layer by layer fashion, training the layer below a layer first and then using the hidden units of the lower layer as visible units for the current layer. The gradient of the log likelihood function of the parameters (W,b,c) can be shown to be

$$\sum_{t=1}^{n} E_{p(h|v^{(t)})}[\nabla_\Theta(-E(v^{(t)}, h))] - n E_{p(h,v)}[\nabla_\theta(-E(v, h))] \tag{4}$$

Where the first term is the expectation under the distribution of data and the second term is the expectation under the model. Computing the above expression can be computationally intensive, but as shown by [11] the expectation under the model can be replaced by a point estimate obtained after k Gibbs sampling. Usually, k equal to 1 is sufficient.

### 2.3 Discriminative fine tuning of DBN using back propagation

Following generative pre training of the DBN, the weights were fine-tuned by constructing a neural network with the same architecture of the DBN. The weights of the neurons were initialized to the weights obtained after pre training. The weights were then fine-tuned using backpropagation on the labelled training examples.

After fine tuning the outputs of the 1st and 2nd layer of the network corresponded to the learnt representations. These are referred to as L1 and L2 features in this paper, respectively.

## 3 Deep belief network for speaker recognition

The objective of the DBN is learning abstract hierarchical representations of (unlabelled) input data. The deep belief network attempts to learn from frame wise spectral data of the speech signal, which is known to carry speaker dependent information (primarily about the vocal tract shape) [12]. Assuming the speaker to be a vibrating body which produces certain spectral features, we believe that deep generative models (like DBN) can be used to recognise and emphasise the patterns within the spectral data which correspond to the characteristic vocal tract shape of the speaker.

The spectrogram (refer Fig. 2) is extracted from the speech data with 25 ms window size with a 10 ms step. To reduce the dimensionality of each frame, PCA whitening is applied on each frame, taking 128 components. We therefore have 128 visible units in the DBN followed by 200 units each in the first and the second hidden layers. Since the inputs to the DBN are real valued, ReLU (Rectified Linear Unit) activation is used for visible units. The activations of the 1st and 2nd hidden layers then form the L1 and L2 features.

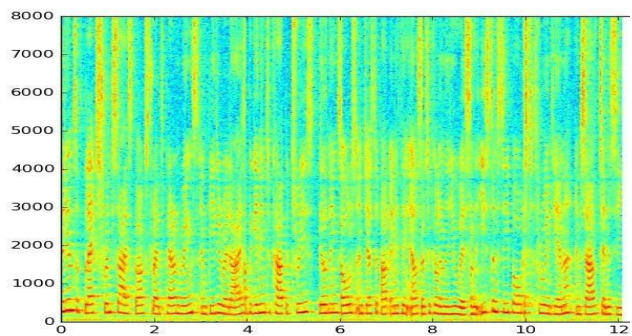

**Fig. 2.** Spectrogram of voice sample, sampled frames of which are used for training the DBN.

We append the features learnt by the DBN for each frame with MFCC features (keeping the 2-14 MFCC coefficients) for that frame and train a GMM-UBM [12] model for classification. This augmented feature vector can better represent the speaker dependent features carried by the spectral frame, than standalone MFCC features.

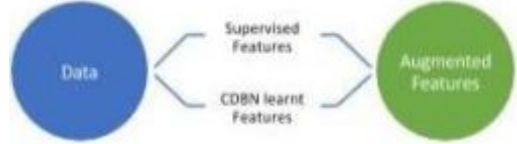

**Fig. 3.** Creation of augmented feature vectors which accurately represent the statistical model of speech.

## 4  Speaker recognition using GMM-UBM framework

For the task of speaker recognition using our learned representations we use the state of the art Gaussian Mixture Model - Universal Background Model framework developed by Reynolds et al. [12]. Gaussian Mixture models [13] are probabilistic models used to represent arbitrary probability distributions. For a GMM with M Gaussians, the probability of a vector x can be expressed as:

$$p(x|\lambda) = \sum_{i=1}^{M} w_i p_i(x))$$
(5)

Where $\lambda = \{w, \mu, \Sigma\}$ refers to the collective parameters for the M gaussians. The probability is the weighted linear sum of M individual Gaussians probabilities $p_i(x)$,

$$p_i(x) = \frac{1}{(2\pi)^{D/2}|\Sigma_i|^{1/2}} \exp\left\{-\frac{1}{2}(x-\mu_i)'(\Sigma_i)^{-1}(x-\mu_i)\right\}.$$
(6)

Where $\mu_i$ is the mean and $\Sigma_i$ is the covariance matrix of the gaussian *i*. Further the sum of all weights $w_i$ is 1.

Reynolds et al.[12] formulated the basic task of speech recognition as a basic hypothesis test. Given a speech segment, *X*, and a hypothesized speaker, *Y*, the test to determine if X was spoken by Y reduces to the hypothesis test,

$H_0 : X$ was spoken by *Y*;    $H_1 : X$ was not spoken by *Y*

They also formulated the optimal test to decide between the two hypotheses as the ratio $\frac{P(X|H_0)}{P(X|H_1)}$. If the ratio is greater than a threshold $\theta$ accept $H_0$ else reject $H_0$. The task is then to model *p(X|H₀)* and *p(X|H₁)*. For each speaker a GMM is fit to augmented features (MFCC + DBN features) with maximum likelihood parameters estimated using the EM [15] algorithm. Each GMM then models the probability of a speech feature being said by a speaker. *p(X|H₁)* is modelled using a single, speaker independent GMM known as the UBM. This GMM is fit with features extracted from speech of all speakers in the dataset. This models the probability of a expected alternative speech which can be encountered. After generating the models for the speaker $P(X|H_0)$ and the background model $P(X|H_1)$, we can use the logarithm of the likelihood:

$$Log\text{-}likelihood = log(p(X|H_0) - log(p(X|H_1)) \qquad (7)$$

The log likelihood can be use as the statistic to verify the speaker. The log likelihood can be normalized using a standard score normalization technique to obtain a confidence measure between 0 and 1.

## 5  Results

We evaluated the performance of our approach on the ELSDSR[15] dataset. We vary the number of speakers and train a GMM-UBM model (with the number of Gaussians as 64 for each GMM). We use different feature vectors for the frames of each sample during training, the results reported are averaged after 15 trials, are as follows:

**Table 1**. Performance of different feature vectors used to train a GMM-UBM model with increase in the number of speakers.

| No of Speakers | MFCC | MFCC + L1 | MFCC + L2 | MFCC + L1 + L2 |
|---|---|---|---|---|
| 2 | 0.98 | 0.99 | 0.98 | 0.99 |
| 5 | 0.94 | 0.97 | 0.95 | 0.98 |
| 12 | 0.92 | 0.96 | 0.91 | 0.96 |
| 15 | 0.90 | 0.95 | 0.91 | 0.96 |
| 22 | 0.90 | 0.94 | 0.90 | 0.95 |

The GMM-UBM model trained using only MFCC features for each frame is beaten by the GMM-UBM models trained by appending the L1 and L2 features obtained from the DBN for the frame.

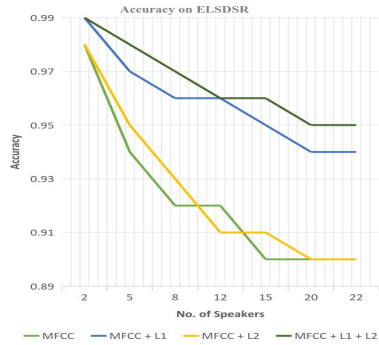

**Fig. 4.** Plot of Accuracy v/s No. of speakers shows superior performance of MFCC + L1+ L2.

We also evaluate our approach by varying the number of utterances per speaker whilst training the GMM. This highlights the dependence of our approach on the availability of labelled data (refer Table 2 and Fig. 5).

**Table 1**. Performance of different feature vectors used to train a GMM-UBM model with increase in the number of utterances per speaker.

| No of Utterances | MFCC | MFCC + L1 | MFCC + L2 | MFCC + L1 + L2 |
|---|---|---|---|---|
| 1 | 0.76 | 0.80 | 0.78 | 0.85 |
| 2 | 0.78 | 0.82 | 0.84 | 0.86 |
| 3 | 0.85 | 0.85 | 0.84 | 0.92 |

This highlights the superiority of our approach in situations where there is lack of labelled data.

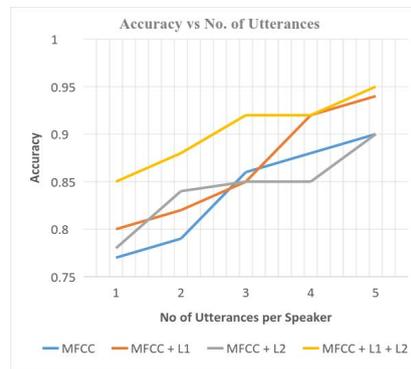

**Fig. 5.** Plot of Accuracy v/s No. of utterances per speaker for 22 speakers shows the superiority of MFCC + L1 + L2 feature vectors.

# 7   Conclusion

Our work therefore highlights the potency of deep belief networks in extracting short term spectral features which are capable of capturing speaker dependent features. We clearly observe that in dearth of labelled training examples, the presence of features learned using the DBN (which can be trained with unlabeled data) greatly enhances the accuracy of speaker recognition in such settings.

Future work can be done with the attempt to find more accurate representation of short term spectral data, to completely remove the use of engineered features such as MFCC.